\documentclass{PoS}
\usepackage{graphicx}
\usepackage{amsmath,amssymb,latexsym}
\usepackage{url}
\usepackage{bm}
\usepackage{xspace}
\usepackage{nicefrac}
\usepackage{epsfig}

\title{Unmasking the ultrahigh-energy cosmic ray origin}

\ShortTitle{Unmasking the ultrahigh-energy cosmic ray origin}

\author{\speaker{Luis A. Anchordoqui}\\
Department of Physics and Astronomy, Lehman College at CUNY, NY 10468,
USA \\
Department of Physics, Graduate Center, City University of New York, NY 10016, USA\\
Department of Astrophysics, American Museum of Natural History, NY 10024, USA\\
E-mail: \email{luis.anchordoqui@gmail.com}}

\abstract{The sharp change in slope of the ultrahigh-energy cosmic ray 
spectrum around $10^{9.6}~{\rm GeV}$ (the ankle), combined with
evidence of a light but extragalactic component near and below the
ankle which evolves to intermediate/heavy composition above, has proved
exceedingly challenging to understand theoretically. Recently, we
introduced a very general model in which, for a range of source conditions,
photo-disintegration of ultrahigh-energy nuclei in the region
surrounding the accelerator naturally accounts for the observed
spectrum and composition of the entire extragalactic component, which
dominates above about $10^{8.5}~{\rm GeV}$.  In this communication we 
review the generalities of the model and show that  starburst galaxies provide a compelling source example.}

\FullConference{EPS Conference on High Energy Physics,\\
		5 - 12 July, 2017\\
		Venice, Italy}



\begin{document}

The ultrahigh-energy ($E > 10^{9}~{\rm GeV}$) cosmic ray (UHECR)
spectrum can be roughly described by a twice-broken power
law~\cite{Abbasi:2007sv}.  The first break, at an energy $E \sim
10^{9.6}~{\rm GeV}$, is a hardening of the spectrum known as the {\it
  ankle}, while the second abrupt softening of the spectrum is the UV cutoff at $E
\sim 10^{10.6}~{\rm GeV}$.  Optical observations
of air showers with fluorescence telescopes or non-imaging Cherenkov
detectors consistently find a predominantly light composition at
around $10^9~{\rm GeV}$~\cite{Kampert:2012mx} and the contribution of
protons to the overall cosmic ray flux is $\gtrsim~50\%$ in this
energy range~\cite{Aab:2014kda}.
Due the absence of a large anisotropy in the arrival direction of
cosmic rays below the
ankle~\cite{Auger:2012an}, we can conclude that
these protons must be of extragalactic origin.  At energies above
$10^{10}~{\rm GeV}$, the high-statistics data from the Pierre Auger
Observatory suggests a gradual increase of the fraction of heavy
nuclei in the cosmic ray flux~\cite{Aab:2014kda}. Within uncertainties, the data
from the Telescope Array (TA) is consistent with these
findings~\cite{Abbasi:2014sfa}.  In addition, TA has
observed a statistically significant excess in cosmic rays with
energies above $10^{10.7}~{\rm GeV}$ in a region of the sky spanning
about $20^\circ$, centered on equatorial coordinates (R.A. =
$146.7^\circ$, Dec. = $43.2^\circ$)~\cite{Abbasi:2014lda}. This is
colloquially referred to as the TA hot spot. The absence of a
concentration of nearby sources in this region of the sky corroborates
other experimental evidence for heavy nuclei, whereby a few local
sources within  our cosmic backyard can produce the hot spot through
deflection in the extragalactic and Galactic magnetic fields.

Simultaneously reproducing the Auger spectrum above the ankle together
with the observed nuclear composition requires hard source
spectra~\cite{Aab:2016zth}. It is possible to accommodate the entire
UHECR data with the addition of an {\it ad hoc} light extragalactic
component below the ankle, with a steep injection
spectrum~\cite{Aloisio:2013hya}. However, a more {\it natural}
explanation of the entire spectrum and composition emerges while
accounting for the {\it post-processing} of UHECRs through
photo-disintegration in the environment surrounding the
source~\cite{Unger:2015laa}.  In this model relativistic nuclei
accelerated by a central engine to extremely high energies remain
trapped in the turbulent magnetic field of the source
environment. Their escape time decreases faster than the interaction
time with increasing energy, so that only the highest energy nuclei
can escape the source unscathed. In effect, the source environment
acts as a high-pass filter on the spectrum of cosmic rays. All nuclei
below the energy filter interact, scattering off the far-infrared
photons in the source environment.  These photonuclear interactions
produce a steep spectrum of secondary nucleons, which is overtaken by
the harder spectrum of the surviving nucleus fragments above about
$10^{9.6}~{\rm GeV}$. These overlapping spectra could then carve an
ankle-like feature into the source emission spectrum. The spectrum
above the ankle exhibits a progressive transition to heavy nuclei, as
the escape of non-interacting nuclei becomes efficient. The model is
essentially characterized by two parameters of the source environment:
{\it (i)}~a low gas density and {\it (ii)}~a thermal background of
far- and mid-infrared photons. In addition, reproducing the data with such a
model requires a central engine with a hard \mbox{spectrum $\propto E^{-1}$,}
as expected for acceleration in young neutron
stars~\cite{Blasi:2000xm}. Note that we can distinguish two
components contributing to the nucleons populating the spectrum below
the ankle: those emitted by the source ${\cal N}_s$ and those produced
during propagation ${\cal N}_p$.

For a central engine emitting a single nuclear species, the
composition that best fit the Auger measurements between the ankle and
the cutoff is $^{28}$Si; see Figs.~1 and 6
in~\cite{Unger:2015laa}. This is because the energy per nucleon $E_N=
E/A$ just above the cutoff $E \sim 10^{10.8}~{\rm GeV}$ has to be
roughly the energy just below the ankle $\sim 10^{9.2}~{\rm
  GeV}$. Such an energy correlation requires a baryon number $A \sim
30$. The model is also consistent with a composition that follows a
Galactic mixture with 10 elements; see Fig.~9 in~\cite{Unger:2015laa}.
For the case of TA data, the favor composition is heavier and
partially consistent with  pure iron; see Fig. 8
in~\cite{Unger:2015laa}.

The possibilities initially entertained conceptually may be brought to
realization in galaxies undergoing periods of rapid star formation,
the so-called ``starbursts.''  These environments feature strong
infrared emission by dust associated with high levels of interstellar
extinction, strong UV spectra from the Lyman $\alpha$ emission of hot
OB stars, and considerable radio emission produced by recent supernova
remnants.  The central regions of starburst galaxies can be orders of
magnitude brighter than those of normal spiral galaxies. By now it is
well-established that from the central active region a galactic-scale
superwind is driven by the collective effect of supernovae and winds
from massive stars~\cite{Veilleux:2005ia}.  The high supernovae rate
creates a cavity of hot gas ($\sim 10^8~{\rm K}$) whose cooling time
is much greater than the expansion time scale. In other words, the
wind has a sufficiently low density that it will move out before
cooling.  Since the wind is sufficiently powerful, it can blow out of
the interstellar medium of the galaxy as a hot bubble.  As the cavity
expands a strong shock front is formed on the contact surface with the
cool interstellar medium.  Observations of local starburst superwinds
seem to indicate that they are a complex multiphase phenomena.  When
the cooler material of interstellar clouds encounters the wind, it is
heated and accelerated, giving rise to regions of optical line
emission. The dust contained in these clouds is revealed as it reddens
the background starlight and scatters the starburst's UV
radiation. Warm ($100 \lesssim T/{\rm K} \lesssim 200$) dust in the
outer layers of the clouds radiates substantially at mid-infrared
wavelength ($530 \lesssim \lambda/\mu{\rm m} \lesssim 300$), and the
cooler dust heated to $30 \lesssim T/{\rm K} \lesssim 60$ radiates in
the far-infrared range ($30 \lesssim \lambda/\mu{\rm m} \lesssim
300$). All in all, the superwind environment satisfies the two
essential requirements advocated in~\cite{Unger:2015laa}. Moreover,
because of the high prevalence of supernovae, starbursts must possess
a large density of newly-born neutron stars.  As noted above, neutrons
stars (with their metal-rich surfaces) are potentially candidate
sources for the central engine: iron nuclei can be stripped off the
surface and be accelerated through unipolar induction up to the
maximum observed energies, and with the appropriate hard spectrum
$\propto E^{-1}$~\cite{Blasi:2000xm}. For surface temperatures
exceeding $10^6$K, photo-disintegration in the thermal radiation
fields generated by the star starts to be effective and with
increasing temperature a large fraction of the initial iron nuclei is
transformed into a mixed composition of lighter
nuclei~\cite{Kotera:2015pya}, akin to that shown in Fig.~9
of~\cite{Unger:2015laa}.

An interesting twist is that after the nuclei escape from the central
region of the galaxy and they are injected into the galactic-scale
superwind they could potentially experience further acceleration at
its terminal shock. The superwind environment could play the role of a
high pass filter, but at the same time could allow re-acceleration of
primary and secondary particles. The shock velocity $v_{\rm sh}$ can
be estimated from the empirically determined superwind kinetic energy
flux $\dot E_{\rm sw}$ and the mass flux $\dot M$ generated by the
starburst through $\dot E_{\rm sw} = \dot M \ v_{\rm sh}^2/2$.  The
shock radius can be approximated by $r_{\rm sh} \approx v_{\rm sh}
\tau$, where $\tau$ is the starburst age. Since the age is about a few
tens to hundreds of million years, the maximum energy attainable in
this configuration is constrained by the limited acceleration time
provided by the shock's finite lifetime. In terms of parameters that
can be determined from observations, for a nucleus of charge $Ze$, we have $E \sim \ Ze \ \overline
B \ \tau \dot E_{\rm sw}/(2\dot M)$, where $\overline B$ is the average
magnetic field~\cite{Anchordoqui:1999cu}. The predicted kinetic energy
and mass fluxes of starbursts, derived from the measured IR
luminosity, are $2 \times 10^{42}~{\rm erg \ s^{-1}}$ and $1.2
M_\odot~{\rm yr}^{-1}$, respectively~\cite{Heckman:1990fe}. The
starburst age is estimated from numerical models that use theoretical
evolutionary tracks for individual stars and make sums over the entire
stellar population at each time in order to produce the galaxy
luminosity as a function of time~\cite{Rieke:1980xt}. Fitting the
observational data these models provide a range of suitable ages for
the starburst phase, $50 \lesssim \tau/{\rm Myr} \lesssim 200 $. It has been
noted that the magnetic energy density of the starburst region could
be significantly higher than that expected from equipartition
arguments with comparable cosmic rays and magnetic energy
densities~\cite{Thompson:2006is}.  Taking $\tau = 50~{\rm Myr}$ and
$\overline B \sim 50~\mu{\rm G}$,  we obtain $E_{N}^{\rm max} \sim
10^{9.5}~{\rm GeV}$. The  source emission spectrum would remain hard
provided its shape is driven by UHECR nucleus leakage from the
boundaries of the shock (a.k.a direct escape)~\cite{Baerwald:2013pu}.

In closing we note that the more involved model, which allows for
re-acceleration of the photo-disintegrated nuclei and the secondary
nucleons, could relax the constraint on nuclear composition by
potentially increasing the ${\cal N}_s/{\cal N}_p$ ratio. Note the
difference between trapped re-accelerated particles and those that
have escaped the source. Future data from AugerPrime and NASA's Probe
Of Multi-Messenger Astrophysics (POEMMA) will provide a unique
opportunity to test the model.

I acknowledge many useful discussions with my 
colleagues of the Pierre Auger Collaboration.  LAA is supported by 
U.S. NSF (Grant No. PHY-1620661) and NASA (Grant No. NNX13AH52G).


\begin{thebibliography}{99}



\bibitem{Abbasi:2007sv} 
  R.~U.~Abbasi {\it et al.} [HiRes Collaboration],
  {\it Phys.\ Rev.\ Lett.}\  {\bf 100}, 101101 (2008);
  J.~Abraham {\it et al.} [Pierre Auger Collaboration],
  {\it Phys.\ Rev.\ Lett.}\  {\bf 101}, 061101 (2008);
  {\it Phys.\ Lett.\ B} {\bf 685}, 239 (2010).
 



\bibitem{Kampert:2012mx} 
  K.~H.~Kampert and M.~Unger,
  {\it Astropart.\ Phys.}\  {\bf 35}, 660 (2012).
 


\bibitem{Aab:2014kda} 
  A.~Aab {\it et al.} [Pierre Auger Collaboration],
  {\it Phys.\ Rev.\ D} {\bf 90},  122005 (2014);
  {\it Phys.\ Rev.\ D} {\bf 90},  122006 (2014);
  {\it Phys.\ Lett.\ B} {\bf 762}, 288 (2016);
  {\it JCAP} {\bf 1704},  038 (2017).



\bibitem{Auger:2012an} 
  P.~Abreu {\it et al.} [Pierre Auger Collaboration],
  {\it Astrophys.\ J.\ Suppl.}\  {\bf 203}, 34 (2012);\\
 A.~Aab {\it et al.} [Pierre Auger Collaboration],
 {\it  Astrophys.\ J.}\  {\bf 802},  111 (2015).
 


\bibitem{Abbasi:2014sfa} 
  R.~U.~Abbasi {\it et al.},
 {\it Astropart.\ Phys.}\  {\bf 64}, 49 (2014);
  R.~Abbasi {\it et al.} [Pierre Auger and Telescope Array Collaborations],
   {\it JPS Conf.\ Proc.}\  {\bf 9}, 010016 (2016).
 

\bibitem{Abbasi:2014lda} 
  R.~U.~Abbasi {\it et al.} [Telescope Array Collaboration],
 {\it  Astrophys.\ J.}\  {\bf 790}, L21 (2014).

\bibitem{Aab:2016zth} 
  A.~Aab {\it et al.} [Pierre Auger Collaboration],
{\it  JCAP} {\bf 1704}, 038 (2017).



\bibitem{Aloisio:2013hya} 
  R.~Aloisio, V.~Berezinsky and P.~Blasi,
{\it   JCAP} {\bf 1410},  020 (2014).

  
\bibitem{Unger:2015laa} 
  M.~Unger, G.~R.~Farrar and L.~A.~Anchordoqui,
  {\it Phys.\ Rev.\ D} {\bf 92}, 123001 (2015).




\bibitem{Veilleux:2005ia} 
  S.~Veilleux, G.~Cecil and J.~Bland-Hawthorn,
  {\it Ann.\ Rev.\ Astron.\ Astrophys.}\  {\bf 43}, 769 (2005)



\bibitem{Blasi:2000xm}
  P.~Blasi, R.~I.~Epstein and A.~V.~Olinto,  
{\it Astrophys.\ J.\ }  {\bf 533}, L123 (2000);\\
  K.~Fang, K.~Kotera and A.~V.~Olinto,
  {\it Astrophys.\ J.\ } {\bf 750}, 118 (2012);
  {\it JCAP} {\bf 1303}, 010 (2013).
 
\bibitem{Kotera:2015pya} 
  K.~Kotera, E.~Amato and P.~Blasi,
 {\it  JCAP} {\bf 1508}, no. 08, 026 (2015).




\bibitem{Anchordoqui:1999cu} 
  L. A. Anchordoqui, G. E. Romero and J. A. Combi,
{\it   Phys.\ Rev.\ D} {\bf 60}, 103001 (1999); \\
  D.~F.~Torres and L.~A.~Anchordoqui,
 {\it Rept.\ Prog.\ Phys.}\  {\bf 67}, 1663 (2004); \\
  L.~A.~Anchordoqui, V.~Barger and T.~J.~Weiler,
  arXiv:1707.05408.



\bibitem{Heckman:1990fe} 
  T.~M.~Heckman, L.~Armus and G.~K.~Miley,
 {\it  Astrophys.\ J.\ Suppl.}\  {\bf 74}, 833 (1990).


\bibitem{Rieke:1980xt} 
  G.~H.~Rieke, M.~J.~Lebofsky, R. Thompson, F. Low and A. Tokunaga,
 {\it  Astrophys.\ J.}\  {\bf 238}, 24 (1980).


\bibitem{Thompson:2006is}
  T. A. Thompson, E. Quataert, E. Waxman, N. Murray and C.  Martin,
  {\it Astrophys.  J. }  {\bf 645}, 186 (2006).
 

\bibitem{Baerwald:2013pu} 
  P.~Baerwald, M.~Bustamante and W.~Winter,
  {\it Astrophys.\ J.}\  {\bf 768}, 186 (2013).




\end{thebibliography}
\end{document}